\documentclass[aps, groupedaddress, amsmath, amssymb, twocolumn,showpacs, preprintnumbers]{revtex4}
\usepackage{graphicx}
\usepackage{dcolumn}
\usepackage{bm}

\begin{document}

\title{Characteristic temperatures of exchange biased systems}
\author{Alexey N. Dobrynin}
     \email{dobrynin@ameslab.gov}
\author{Ruslan Prozorov}
\affiliation{Ames Laboratory and Department of Physics \& Astronomy, Iowa State University,
Ames, IA 50011}

\begin{abstract}
Characteristic temperatures in ferromagnetic - antiferromagnetic exchange biased systems are analyzed.
In addition to usual blocking temperature of exchange bias $T_{B}$, and
the N\'{e}el temperature of an antiferromagnet $T_{N}$, the
inducing temperature $T_{ind}$, i.e., the temperature, at which the
direction of exchange anisotropy is established, has been recently proposed. We demonstrate that this temperature is in
general case different from $T_{B}$ and $T_{N}$. Physics and experimental
approaches to measure the inducing temperature are discussed. Measurements of $T_{ind}$, in addition to $T_{B}$, and $T_{N}$, 
provide important information about exchange interactions in ferromagnetic -
antiferromagnetic heterostructures.
\end{abstract}

\pacs{75.30.Gw, 75.70.Cn, 75.30.Et, 75.50.Ee}
\maketitle

\indent Exchange anisotropy appears in hybrid ferromagnetic (F) -
antiferromagnetic (AF) systems due to exchange interactions at the F-AF
interface~\cite{MB}. The interfacial exchange creates an additional energy
barrier, which F magnetic moments have to overcome during the magnetization
reversal. The exchange anisotropy is unidirectional and shows up as a
horizontal shift of the magnetic hysteresis loop after field cooling, and the
exchange bias field is determined as the value to which the center of the
hysteresis loop is shifted with respect to the zero field~\cite{Nogues}.
This assumes that the AF structure stays stable, which is valid unless the
total AF magnetocrystalline anisotropy is too low, when AF spins rotate
coherently with F spins, and the exchange bias vanishes~\cite{Meik62, Lund,
Dobrynin}. The loop is normally shifted in the direction opposite to the
cooling field, which indicates that the interfacial exchange coupling is
ferromagnetic, i.e., it favors parallel orientation of the interfacial F and
AF spins. The case of \textquotedblleft positive\textquotedblright\ loop
shifts, which may assume antiferromagnetic coupling at the F-AF interface
(favoring antiparallel alignment of the interfacial F and AF spins), was
described by Nogu\'{e}s et al.~\cite{NogPEB, NogAFcoupl}. \newline
\indent It is a \textquotedblleft common knowledge\textquotedblright\ that
the exchange anisotropy is established when field cooling a F-AF system
through the N\'{e}el temperature $T_{N}$ of the antiferromagnet~\cite{Nog051, Berkowitz}. The blocking temperature of exchange bias $T_{B}$ is the temperature, at which exchange bias disappears. It has been recently
demonstrated that the direction of exchange anisotropy can be established at
a temperature larger than $T_{B}$, which is determined as exchange bias
inducing temperature $T_{ind}$~\cite{Dobrynin3}.\newline
\begin{center}
\begin{figure}[tbp]
\centering
\includegraphics [width=8.0792cm] {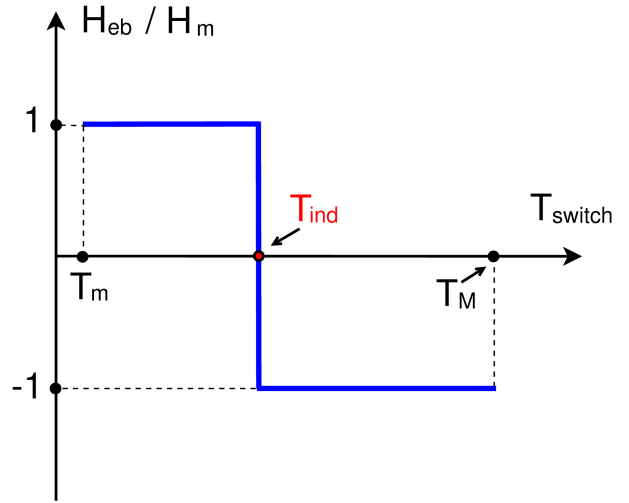}
\caption{\label{cooling} Fig.~1.  Exchange bias field $H_{eb}$, measured at temperature $T_m$ as a function of temperature $T_{switch}$, at which the direction of the cooling field is changed from $-H_{FC}$ to $+H_{FC}$. The temperature is scanning from $T_M$ down to $T_m$. In case of $T_{ind} < T_{switch} < T_M$ the exchange bias is negative ($-H_m$), since it's induced by a positive cooling field $+H_{FC}$. For $T_m < T_{switch} < T_{ind}$ the exchange bias is positive ($+H_m$), because it's induced by a negative cooling field $-H_{FC}$. $T_{ind}$ is the temperature, at which the direction of the exchange anisotropy is established.}
\end{figure}
\end{center}
\indent The procedure of measuring $T_{ind}$ is as follows. At first the
sample is field cooled in a \textquotedblleft negative\textquotedblright\
field $-H_{FC}$ from temperature $T_{M}$ ($T_{M}>T_{N}$), to a certain
temperature $T_{switch}$, where the sign of the cooling field is changing.
The further cooling to the temperature $T_{m}$ is performed at field $+H_{FC}
$. $T_{m}$ is the temperature, at which the hysteresis loop is measured ($T_{m}<T_{B}$ should be satisfied). The absolute value of the exchange bias field at $T_{m}$ is $H_{m}$. If the direction of the exchange anisotropy is not established at $T_{switch}$, then the exchange bias field, measured at $T_{m}$ will be $-H_{m}$, since it will be induced in a positive cooling field $+H_{FC}$. In the second case, the direction of exchange anisotropy will be established at a temperature higher than $T_{switch}$, and changing the sign of the cooling field does not influence the sign of the exchange bias field $+H_{m}$, measured at $T_{m}$. By scanning $T_{switch}$ from $T_{M}$ down to $T_{m}$, the transition temperature $T_{ind}$ will be found, at which the direction of exchange anisotropy is established~\cite{comment1}. The dependence of the exchange bias field $H_{eb}$, measured at $T_{m}$, versus $T_{switch}$, is schematically illustrated in Fig.~1. In the above description we assume that there is no training effect in the system~\cite{Hoffmann}. Otherwise, the transition at $T_{ind}$ would be not from $-H_{m}$ to $+H_{m}$, but from $-H_{m}$ to $H_{mTL}$, where the last value is the exchange bias field of the first training loop at $T_{m}$. \newline

\indent In order to understand the origin of the inducing temperature, and
its difference from the blocking temperature, we consider a one-dimensional
F-AF model system, which is schematically shown in Fig.~2. In
the assumption that the exchange interactions exist only between localized
nearest-neighbor spin magnetic moments, the exchange Hamiltonian of the
system may be written as:
\begin{eqnarray}  \label{Hex}
H_{ex} &=& - J_F\sum_{i=1}^{N_F-1} \mathbf{S_F}_i\mathbf{S_F}_{i+1} -
J_A\sum_{j=1}^{N_A-1} \mathbf{S_A}_j\mathbf{S_A}_{j+1} - {}  \nonumber \\
& & {} - J_{int}\mathbf{S_F}_{N_F}\mathbf{S_A}_{N_A}
\end{eqnarray}
Here $\mathbf{S_F}$ and $\mathbf{S_A}$ are F and AF spin magnetic moments
respectively, $J_F > 0$ is the exchange coupling constant between F spins, $J_A < 0$ is the exchange coupling constant between AF spins, and $J_{int} > 0$ is the exchange coupling constant between interfacial F and AF spins $\mathbf{S_F}_{N_F}$ and $\mathbf{S_A}_{N_A}$ respectively.\newline

\begin{center}
\begin{figure}[tbp]
\centering
\includegraphics [scale=0.9] {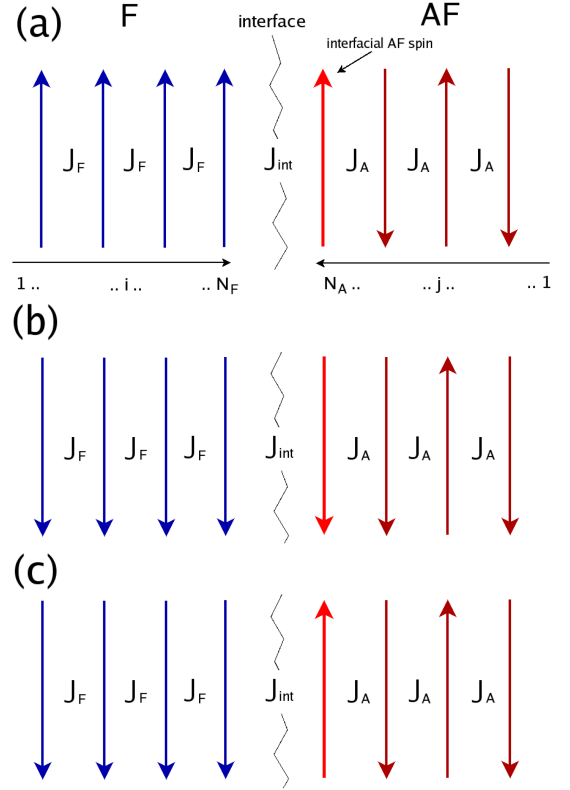}
\caption{\label{interface} Fig.~2. Schematic view of the one-dimensional F-AF system. (a) The system is cooled in a ``positive'' field, and the ground state is formed. (b) $J_{int} > J_A$: the interfacial AF spin rotates coherently with the F part, and the exchange bias field is determined by $J_A$. (c) $J_{int} < J_A$: the interfacial AF spin stays stable during the magnetization reversal, and the exchange bias field is determined by $J_{int}$.}
\end{figure}
\end{center}
\indent  The first term in Eq.~\ref{Hex} is responsible for the stability of
the F magnetic structure, the second term - for that of the AF structure,
and the third one - for the interfacial F-AF coupling, which causes the
exchange anisotropy. The direction of the exchange anisotropy is determined
by the orientation of $\mathbf{S_A}_{N_A}$, since in the ground state F
spins are parallel to the interfacial AF spin, as imaged in Fig.~2~(a). We define ${T_N}_{int}$, as the temperature, at which the AF exchange interaction between  $\mathbf{S_A}_{N_A}$ and  $\mathbf{S_A}_{N_A-1}$ is established. The temperature, at which the interfacial F-AF interaction (i.e., interaction between $\mathbf{S_F}_{N_F}$ and $\mathbf{S_A}_{N_A}$) is established, is designated as $T_{FAF}$. ${T_N}_{int}$ is proportional to $J_A$, while $T_{FAF}$ is proportional to $J_F$ (in many dimensional case these freeing temperatures are proportional to the product of the corresponding exchange coupling constant and the corresponding coordination number).  \newline
Assume that the system was cooled in a ``positive'' field through $T_N$, and
the direction of the field was changed just after passing $T_N$. ${T_N}_{int}
$ is less than $T_N$ due to the reduced AF coordination number at the
interface. Therefore, the interfacial spin can still be aligned by the
external negative field, yielding the configuration, shown in Fig.~2~(b). However this is a metastable state, rather than the ground state of the system, since $\mathbf{S_A}_{N_A}$ favors antiparallel orientation with $\mathbf{S_A}_{N_A-1}$. The ground state still will be that, shown in Fig.~2~(a), and therefore, the direction of exchange anisotropy is established when passing the $T_N$, i.e., $T_{ind}=T_N$ for the one-dimensional case.\newline

\indent In the real three-dimensional system the above described situation
is not the only possibility. Most of the models of exchange bias assume a
kind of frustrated interfacial AF spin configuration \cite{Leigh-PRL}. In
particular, uncompensated interfacial AF spins were found to be the reason
of exchange bias in many systems \cite{Takano, Hoff, Ohldag, Kapp}. Such an
uncompensated AF spin has both parallel and antiparallel AF neighbors, and,
therefore, is in a frustrated state. Similar configurations may exist if
other mechanisms are involved in the exchange anisotropy, such as spin-flop
coupling \cite{Koon, Ijiri}, hybrid F-AF domain walls \cite{Chien} or
partial AF domains \cite{Krivorotov}. The frustrated state of the
interfacial AF spins in combination with the reduced AF coordination number
at the interface leads to the situation, when ${T_N}_{int}$ is less than $T_N$, and the interfacial AF spins can be reoriented by an external field above this temperature, and below $T_N$ \cite{Leighton-2}. When further field cooled, the frustrated interfacial AF spin will couple to the neighboring AF spin with antiparallel orientation, forming a ground state. This direction will be the easy direction of magnetization of the whole system. This way, the temperature at which the most favorable orientation of the interfacial AF spins is established, is the inducing temperature of exchange bias.\newline
\indent While the fact of the difference of $T_B$ and $T_N$ is well
established \cite{Zaag}, the strict definition of $T_B$ is missing. It is a
common way to determine $T_B$ as the maximal temperature at which exchange
bias exists after field cooling a F-AF system through $T_N$. We find it
appropriate to accept this as a definition of $T_B$. This way, $T_B$ and $T_{ind}$ are easily measurable values, which in different situations correspond to real freezing temperatures $T_N$ (also measurable), $T_{FAF}$, or ${T_N}_{int}$. Below we discuss these possibilities.\newline
\indent Obviously, exchange bias can not exist while all interfacial exchange interactions are established. Thus, $T_B = min (T_{FAF}, {T_N}_{int})$. If the interfacial F-AF exchange energy is weaker than the exchange energy between the interfacial AF spins and the rest of the AF part ($J_{int} < J_A$ for the one dimensional case), then the measurements of $T_B$ will yield $T_{FAF}$, while $T_{ind}$ corresponds to ${T_N}_{int}$ (frustrated case) or $T_N$ (non-frustrated case), and, therefore, $T_B < T_{ind}$. This also means that the interfacial AF spins will stay stable during the magnetization reversal at $T_m$, and the exchange bias value is determined by $J_{int}$. This situation is shown in Fig.~2~(c).\newline
If  $J_{int} > J_A$ then  $T_{FAF} > {T_N}_{int}$. The interfacial uncompensated (i.e., those, responsible for exchange bias) AF spins will rotate coherently with the F spins during the magnetization reversal at $T_m$, and the exchange bias value is determined by $J_A$. If the interfacial AF structure is frustrated, then the measurements of $T_B$ will yield ${T_N}_{int}$, as well as the measurements of $T_{ind}$. If $J_{int} > J_A$, and the interfacial AF structure is not frustrated, $T_B$ will be less than $T_{ind}=T_N$, because it still corresponds to ${T_N}_{int}$, while the direction of exchange bias will be set at $T_N$, as was discussed for the one-dimensional case. This situation corresponds to one, shown in Fig.~2~(b).\newline
\indent A slight difference between $T_B$ and $T_{ind}$ has been recently
observed in oxidized Co nanocluster films \cite{Dobrynin3}. This difference
is expected to be larger for systems with rough F-AF interface, where $J_{int}$ is significantly reduced as compared to that in natively oxidized or epitaxially grown F-AF systems \cite{NogRough, Leigh-PRB}.\newline
\indent Sometimes the technique, developed by Soeya et al.~\cite{Soeya}, is
used to determine $T_B$. With this method a sample is first field cooled to
the temperature $T_m$, then warmed up at zero field to a certain
temperature, at which the magnetic field of the opposite sign is applied,
and the sample is cooled back to $T_m$ at this field. The temperature, at
which the direction of exchange bias, measured at $T_m$, can be changed, is
accepted as $T_B$. Apparently, this technique will yield the same result, as
the technique for measuring $T_{ind}$ unless there is some thermal
hysteresis in the system. Thus, Soeya et al. method will not yield the
 true $T_B$ value in all situations, as discussed above. \newline

\indent In this letter we have demonstrated that it is necessary to
distinguish between the temperature at which the direction of exchange
anisotropy is established ($T_{ind}$), the maximal temperature, at which
exchange bias may exist ($T_B$), and the N\'eel temperature of the
antiferromagnet ($T_N$) in F-AF heterostructures. The modified method of
measuring $T_{ind}$ was proposed, and the method, yielding the true $T_B$
value has been highlighted. Moreover, important information about
interfacial F-AF structure and exchange interactions may be extracted by
comparing these three temperatures. The case of $T_{ind} < T_N$ suggests
presence of a frustrated interfacial AF structure in a system, otherwise $T_{ind}=T_N$. If $T_B = T_{ind} < T_N$, the interfacial F-AF interactions are stronger than that between the interfacial AF spins and the rest of the AF part, assuming rotation of the interfacial AF spins during the magnetization reversal. The exchange bias value in this case is determined by the latter AF exchange coupling. In the case of $T_B < T_{ind} < T_N$ the interfacial AF spins stay stable, and the exchange bias field is determined by the interfacial F-AF exchange coupling. Systematic comparison of $T_{ind}$, $T_B$, and $T_N$ in different exchange biased systems will help to reveal
the involved exchange mechanisms, and understand better the exchange bias phenomenon. \newline

\begin{acknowledgments}
A. D. appreciates  discussions with P. Lievens, K. Temst, and J. Nogu\'es. Ames Laboratory is operated for the U.S. Department of Energy by Iowa State University under Contract No. W-7405-ENG-82. This work was supported in part by the Director for Energy Research, Office of Basic Energy Sciences. R. P. acknowledges support from the Alfred P. Sloan Foundation.
\end{acknowledgments}


\begin{thebibliography}{26}
\bibitem[{com()}]{comment1}
\bibinfo{note}{The described technique is slightly different from that,
  proposed in Ref.~\cite{Dobrynin3}. In the latter case different values of
  positive and negative cooling fields were used (external field cooling and
  remanent field cooling). This leads to different values of the exchange bias
  field at $T_m$. While not affecting the final result, this aproach introduces
  unnecessary complications to the $T_{ind}$ measuring procedure.}
\end{thebibliography}

\end{document}